\documentclass[superscriptaddress,showpacs,preprintnumbers,amsmath,amssymb,twocolumn]{revtex4-1}
\usepackage{graphicx}
\usepackage{verbatim} %needed for multiline comments
\usepackage{color}
\usepackage[colorlinks]{hyperref}
\usepackage{epstopdf}

\usepackage{dcolumn}% Align table columns on decimal point
\usepackage{bm}% bold math
\usepackage{upgreek}

\newcommand{\erfc}[1]{\mathop{\rm erfc}{#1}}

\usepackage[normalem]{ulem}

\begin{document}

\title{
Microscopic dynamics of electron hopping in a semiconductor quantum well probed by spin-dependent photon echoes\\
%Spin transport in hopping regime via photon echo\\
%Local/Microscopic spin transport revealed by photon echoes\\
%Photon echoes as a probe of local spin transport in quantum wells\\
%Electron transport via spin-dependent photon echo
}

\author{A.~N.~Kosarev}
 	\affiliation{Experimentelle Physik 2, Technische Universit\"at Dortmund, 44221 Dortmund, Germany}
 	\affiliation{Ioffe Institute, 194021 St. Petersburg, Russia}
 	\author{S.~V.~Poltavtsev}
 	\affiliation{Experimentelle Physik 2, Technische Universit\"at Dortmund, 44221 Dortmund, Germany}
 	\affiliation{Spin Optics Laboratory, St. Petersburg State University, 198504 St. Petersburg, Russia}
\author{L.~E.~Golub}
\author{M.~M.~Glazov}
 	\affiliation{Ioffe Institute, 194021 St. Petersburg, Russia}
\author{M.~Salewski}
\affiliation{Experimentelle Physik 2, Technische Universit\"at Dortmund, 44221 Dortmund, Germany}
\author{N.~V.~Kozyrev}
 	\affiliation{Ioffe Institute, 194021 St. Petersburg, Russia}
\author{E.~A.~Zhukov}
\author{D.~R.~Yakovlev}
 	\affiliation{Experimentelle Physik 2, Technische Universit\"at Dortmund, 44221 Dortmund, Germany}
 	\affiliation{Ioffe Institute,  194021 St. Petersburg, Russia}
\author{G.~Karczewski}
\author{S.~Chusnutdinow}
 	\affiliation{Institute of Physics, Polish Academy of Sciences, PL-02668 Warsaw, Poland}
\author{T.~Wojtowicz}
 	\affiliation{International Research Centre MagTop, Institute of Physics, Polish Academy of Sciences, PL-02668 Warsaw, Poland}
\author{I.~A.~Akimov}
\author{M.~Bayer}
 	\affiliation{Experimentelle Physik 2, Technische Universit\"at Dortmund, 44221 Dortmund, Germany}
 	\affiliation{Ioffe Institute, 194021 St. Petersburg, Russia}

\date{\today}

\begin{abstract}
Spin-dependent photon echoes in combination with pump-probe Kerr rotation are used to study the microscopic electron spin transport in a CdTe/(Cd,Mg)Te quantum well in the hopping regime. We demonstrate that independent of the particular spin relaxation mechanism, hopping of resident electrons leads to a shortening of the photon echo decay time, while the transverse spin relaxation time evaluated from pump-probe transients increases due to motional narrowing of spin dynamics in the fluctuating effective magnetic field of the lattice nuclei.
\end{abstract}

\keywords{Larmor oscillations, Kerr rotation, Four-wave-mixing, Coherence and relaxation, Excitons, Trions, Quantum wells, Optics of semiconductors}

\maketitle

Understanding the charge and spin dynamics in condensed matter is essential for the development of novel spintronic devices in which the combination of charge transport with ultrafast spin initialization using optical pulses can be exploited~\cite{Dyakonov-book, Zutic-04, Awschalom-09, Kukushkin-16, Belykh-18}. In semiconductors, conduction band electrons which are localized on donor atoms or potential fluctuations, demonstrate long spin relaxation times due to suppression of spin-orbit effects~\cite{glazov:book}. On the other hand, hopping of electrons between localization sites or spin transfer between electrons due to exchange interaction may become relevant and govern the spin dynamics~\cite{Kavokin-02}. The importance of hopping of  conduction band electrons was manifested in studies on optical orientation and spatial diffusion of spin gratings where it was used to uncover the spin relaxation mechanism of electrons~\cite{Korenev-02,Cundiff-06}. Transport effects such as the spin Hall effect in the hopping conductivity regime may occur in disordered two-dimensional systems which have attracted recently significant attention~\cite{Yu-15, Golub-17,Golub-18}. However, previous studies focused on the macroscopic properties of large ensembles, preventing insight into the local dynamics on the sub-$\mu$m scale. Access to microscopic spin and charge dynamics has remained challenging also because the down-scaling of an experiment down to the level of single carrier spins diminishes the correlations between quasi-particles. Therefore, the development of new approaches for the investigation of the microscopic charge and spin properties in large ensembles are in high demand.

Time-resolved optical techniques allow one to access the spin dynamics of both photoexcited and resident carriers~\cite{Dyakonov-book, Awschalom-book}. Using them, one typically detects the macroscopic polarization of an ensemble of spins with a non-zero ensemble average, which was induced by a circularly polarized laser pulse due to optical orientation of excitonic complexes. The most prominent examples for such techniques are polarized photoluminescence~\cite{OO-book, Amand-02} and pump-probe Faraday/Kerr rotation~\cite{Crooker-97, Yugova2009, Glazov2010}. Recently, a novel technique based on spin-dependent photon echoes was introduced which potentially is well suited to investigate the spin dynamics of resident carriers in semiconductors~\cite{Langer-12, Langer-14, Salewski-17}. The unique feature of photon echoes is the reversal of dephasing processes in an ensemble of emitters with an inhomogeneous broadening of optical transitions~\cite{Hartmann-69}. Thereby, echo techniques provide access not only to the homogeneous linewidth of the optical transition, but also show exceptionally high sensitivity to spectral diffusion, e.g. due to energy relaxation or resonance frequency variation~\cite{Cundiff-91}. However, this approach has not yet been applied to resident carriers so far.

In this letter we demonstrate that using spin-dependent photon echoes in combination with pump-probe Faraday/Kerr rotation we can monitor the local spin dynamics of resident electrons and measure the hopping rate between the localization sites which so far had remained a free parameter in model descriptions. We evaluate hopping times in the order of several to tens of ns at low temperatures for electrons with a low density of $10^{10}$~cm$^{-2}$ in a CdTe/(Cd,Mg)Te quantum well (QW). If  hopping is suppressed, both techniques give the same decay time of the coherent signal, which corresponds to the transverse electron spin relaxation. When the hopping rate, on the other hand, becomes comparable to the spin relaxation rate the photon echo decay is accelerated. By contrast, in pump-probe the decay time increases due to motional narrowing in the fluctuating effective nuclear magnetic field, enhancing the spin coherence. In full accord with the developed theoretical model the hopping rate increases with increasing temperature while it decreases in the limit of stronger localization, e.g., when electrons are bound to donors as compared to electrons localized on potential fluctuations.

\begin{figure}[htp]
\includegraphics[width=0.8\columnwidth]{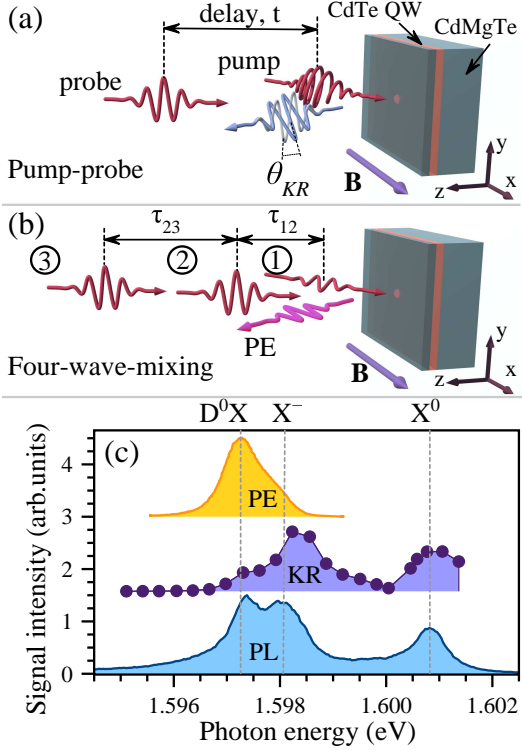}
\caption{Schemes of pump-probe (a) and four-wave-mixing (b) experiments including sketch of the investigated CdTe/(Cd,Mg)Te QW structure. The angle of Kerr rotation $\theta_{KR}$ or ellipticity is used to detect $S_z(t)$. (c) Spectral dependence of photoluminescence intensity and amplitude of long-lived signals measured using Kerr rotation and photon echo. $B=0.25$~T, $T=1.5$~K.}
\label{fig1}
\end{figure}

% Sample 032112B  + description of the main experimental parameters
The studied sample and the scheme of experiments are shown in Figs.~\ref{fig1}(a) and \ref{fig1}(b). The semiconductor structure comprises a $20\,$nm-thick CdTe single QW sandwiched between Cd$_{0.78}$Mg$_{0.22}$Te barriers grown on a (100)-oriented GaAs substrate by molecular-beam epitaxy. The unavoidable background of impurities results in low density resident electrons that are localized both on potential fluctuations and on donors in the QW~\cite{Salewski-17}. This is confirmed by the photoluminescence (PL) spectrum shown in Fig.~\ref{fig1}(c), which shows three optical transitions that are attributed to the neutral exciton (X$^0$) centered at photon energy 1.6008~eV, the negatively charged excitons (trion X$^-$) at 1.5981~eV and the donor-bound excitons (D$^0$X) at 1.5972~eV in accord with Refs.~\cite{Salewski-17,Poltavtsev-17}.  The sample was mounted in the variable temperature insert of a magneto-optical cryostat with the superconducting coil oriented for the Voigt geometry with the magnetic field $\textbf{B}$ normal to the optical axis and parallel to the sample plane ($\textbf{B}\parallel \textbf{x}$). We studied the spin dynamics of resident electrons using time-resolved Kerr rotation (KR) and transient four-wave mixing (FWM). In both experiments resonant excitation of the exciton complexes was obtained with a tunable self-mode-locked Ti-Sapphire laser as source of optical pulses with spectral width of 0.9~meV and duration of $2 - 3\,$ps at a repetition rate of $75.75\,$MHz. The optical pulses applied in each scheme have all the same central photon energy $\hbar\omega$ (degenerate configuration) and hit the sample close to normal incidence with wavevectors $\mathbf{k}_i$, where $i=1,2,3$ is the pulse number in the sequences in Figs.~\ref{fig1}(a) and \ref{fig1}(b). The excitation spot diameter was about 200~$\mu$m and the pulse powers were kept low enough to remain in the linear regime with respect to single pulse excitation. Further details on the KR and FWM experimental setups can be found in Refs.~\onlinecite{Zhukov-07} and \onlinecite{FFT-obzor}.

In the KR experiment a circularly-polarized pump pulse creates a macropscopic spin polarization of resident electrons along the $z$-axis, with $z$ being the sample normal~\cite{Yugova2009,Glazov2010}. In the external magnetic field the electron spins precess in the $yz$-plane with the Larmor precession frequency $\Omega_L=g_e\mu_BB/\hbar$, where $g_e$ is the electron $g$ factor, $\mu_B$ is the Bohr magneton, and $\hbar$ is the reduced Planck constant. The $z$ component of the ensemble averaged spin density, $S_z$, is detected by the spin-Kerr and ellipticity effects, which result in corresponding variations of the polarization of the reflected, linearly-polarized probe beam, see Fig.~\ref{fig1}(a). Scanning the delay time $t$ between the pump and the probe allows us to measure the spin dynamics of the macroscopic spin polarization and to determine both frequency $\Omega_L$ and decay time $T_{\rm 2,KR}^*$ by fitting the experimental data with $S_z(t) \propto \cos{(\Omega_L t)}\exp(-t/T_{\rm 2,KR}^*)$.

\begin{figure}[ht]
\includegraphics[width=0.8\columnwidth]{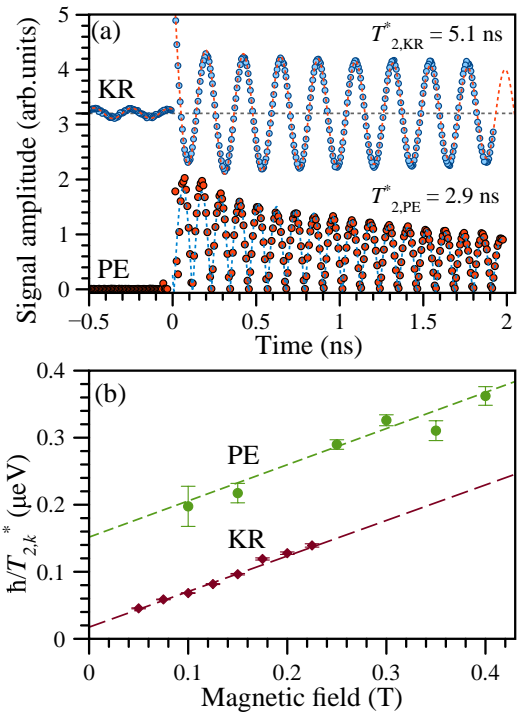}
\caption{(Color online) (a) Examples of pump-probe Kerr ellipticity (labeled KR) and PE transients measured for resonant trion (X$^-$) excitation at 1.5981~eV. $T = 1.5$~K and $B=0.2$~T. The 3-pulse photon echo amplitude is measured for fixed delay $\tau_{12} = 26.7$~ps. Dashed curves show fits using exponentially damped oscillatory function with $\Omega_L=28.1$~rad/ns for both traces, as well as $T^*_{\rm 2,KR}=5.1$~ns, and $T^*_{\rm 2,PE}=2.9$~ns. (b) Magnetic field dependence of relaxation rate for X$^-$ at $1.5981$~eV obtained from the KR and PE signals. $T = 1.5$~K, $\tau_{12} = 26.7$~ps. Results are fitted using linear field dependence (dashed lines) with the same $\Delta g_e = 9\times 10^{-3}$,  as well as $T_{\rm 2, KR} \gtrsim  20$~ns, and $T_{\rm 2,PE} = 4.3 \pm 0.7$~ns.}
\label{fig2}
\end{figure}

In the transient FWM experiment a sequence of three laser pulses is used to generate a coherent optical response along the phase matching direction $\mathbf{k} _{\rm FWM}=\mathbf k_3+ \mathbf k_2-\mathbf k_1$, where in our case $\mathbf k_2= \mathbf k_3$~\cite{Shah-book}. Due to inhomogeneous broadening of the optical transitions this response is given by photon echoes (PEs). Here we focus on the 3-pulse PE, which appears at the delay time $t_{\rm PE}  = 2\tau_{12} + \tau_{23}$ relative to the arrival time of pulse 1, where $\tau_{ij}$ is the delay time between pulses $i$ and $j$. For resonant excitation of X$^-$ or D$^0$X complexes, see Fig.~\ref{fig1}(c), PEs appear even for long delays, decaying on the time scale of several ns when $\tau_{23}$ is scanned, Fig.~\ref{fig2}(a). This decay time is significantly longer than the lifetime of the optical excitations below $ 100$~ps~\cite{Salewski-17}. Here, the pulse 1-2 sequence orients the spins of each resident electron depending on the excitonic resonance frequency $\omega_0$ and the delay time $\tau_{12}$. As a result, a spin grating in coordinate and frequency space is formed
\begin{multline}
\label{grating}
S_y^0 {-\mathrm i S_z^0} \propto
	{\mathrm i \exp\left({\mathrm i\Omega_L\tau_{12}\over 2}\right)
	\cos{\left(\omega_0\tau_{12} + \mathbf k_{2,\parallel}\mathbf r - \mathbf k_{1,\parallel} \mathbf r \right)}}
\end{multline}
for orthogonally linearly polarized pulses $1$ and $2$ with the in-plane components of the wavevectors $\mathbf k_{1,\parallel}$ and $\mathbf k_{2,\parallel}$, respectively. The spin grating is retrieved optically with the third pulse, which induces the 3-pulse PE~\cite{Langer-14}. The time evolution of the grating distribution~\eqref{grating} before the arrival of pulse 3 is characterized by the electron Larmor precession and the decay of the spin grating proportional to $\cos{(\Omega_L t)}\exp(-\tau_{23}/T_{\rm 2,PE}^*)$. In contrast to the established transient spin grating technique~\cite{Awschalom-09, Cundiff-06}, where $\tau_{12}\approx 0$, the spin-dependent PE appears when $\Delta \omega_0 \tau_{12} \gg \pi$ with $\Delta \omega_0$ being the spectral width of the addressed optical transitions. Therefore, the PE signal is very sensitive to spectral diffusion of the resident electrons, which is important in the hopping regime. Variation of the delay time $\tau_{23}$ allows one to extract $\Omega_L$ and $T_{\rm 2,PE}^*$ using the above form for fitting, similar to KR. Note that we measure the absolute value of the electric field amplitude at the PE peak maximum and, therefore, the signal is described by the modulus of the oscillatory function, as shown in Fig.~\ref{fig2}(a)~\cite{FFT-obzor}.

The spectral variation of the signal amplitudes of the resident electron relaxation dynamics are shown in Fig.~\ref{fig1}(c). In contrast to PE, where long-lived signals are observed only for resonant excitation of X$^-$ and D$^0$X, long-lived KR signal is present also for resonant excitation of neutral excitons, because even in a degenerate pump-probe experiment excitation and detection of spin polarization may be performed at different optical transitions. First we concentrate on the resonant excitation of trions, where the differences in relaxation times between the KR and PE results are most pronounced. The KR and PE transients are shown in Fig.~\ref{fig2}(a). Both signals contain additional contributions from the spin dynamics of optically excited carriers with short response times below 100~ps. Here, we concentrate on the long-lived spin dynamics, which is contributed only by resident electrons. One expects to observe the same relaxation behaviour in KR and PE, as the decay is governed by spin dephasing of the electron ensemble, i.e. $T_{\rm 2,PE}^* = T_{\rm 2,KR}^*$. Indeed, the Larmor precession frequencies at $B=0.2$~T are the same, $\Omega_L = 28.1$~rad/ns, in both cases. However, the decay times are surprisingly different. For KR we obtain a notably longer time $T^*_{\rm 2,KR} = 5.1$~ns as compared to the shorter PE decay time $T^*_{\rm 2,PE}=2.9$~ns. Spin dephasing in transverse magnetic field can result from the spread of  electron $g$ factor $\Delta g_e$~\cite{Zhukov-07}. To account for this contribution, we measured the KR and PE signals for different magnetic fields and evaluated the dependences of $\Omega_L$ and dephasing times on $B$. Both methods give identical $B-$linear dependencies of $\Omega_L$, from which we obtain $|g_{e}| = 1.60$, in agreement with previous reports~\cite{Zhukov-07, Saeed-18, Salewski-17}. Details are given in the Appendix A.

The magnetic field dependencies of the decay rates $1/T^*_{\rm 2,PE}$ and $1/T^*_{\rm 2,KR}$ are plotted in Fig.~\ref{fig2}(b). Both show a linear rise with increasing magnetic field and can be described by $1/T^*_{\mathrm 2,k} = 1/T_{\mathrm 2,k} + \Delta g_e \mu_B B/\hbar$ ($k$ is KR or PE),  with the same slope corresponding to $\Delta g_e = 9 \times 10^{-3}$. This confirms that the KR and PE signals are provided by the same sub-ensembles of resident electrons. However, the relaxation times $T_{2,{\rm KR}} \gtrsim 20$~ns and $T_{2,{\rm PE}}\approx 4$~ns are drastically different.

\begin{figure}[ht]
\includegraphics[width=\columnwidth]{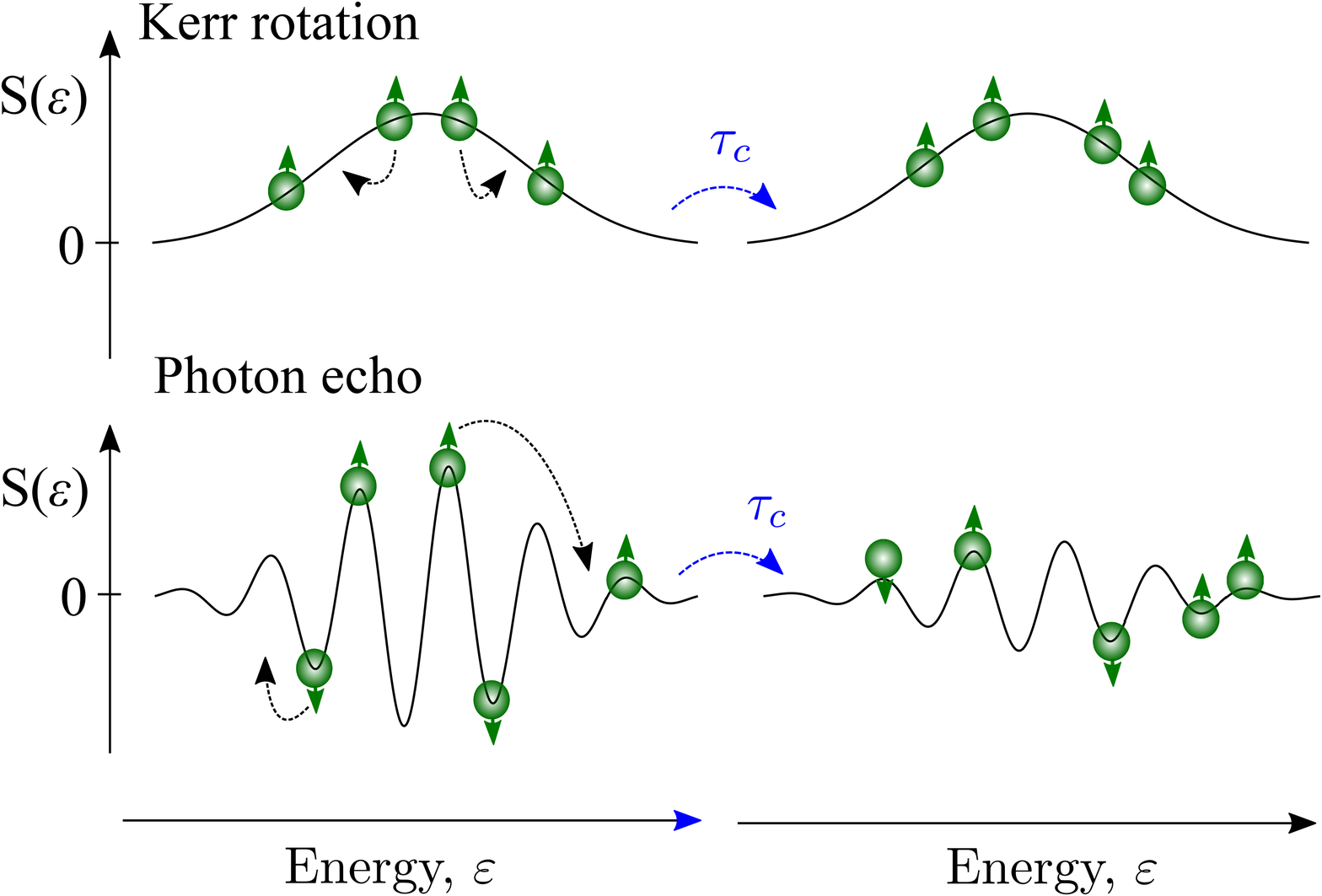}
\caption{Scheme of energy distribution of spin density $S(\varepsilon)$ of localized resident electrons in the KR (top) and PE (bottom) experiments before (left) and after (right) hopping between localization sites. A single hop is accompanied by a change of energy $\varepsilon$, while spin is conserved. Therefore, hopping destroys the spectral grating, but does not influence the macroscopic spin.}
\label{fig3}
\end{figure}

\begin{figure}[htp]
\includegraphics[width=0.85\columnwidth]{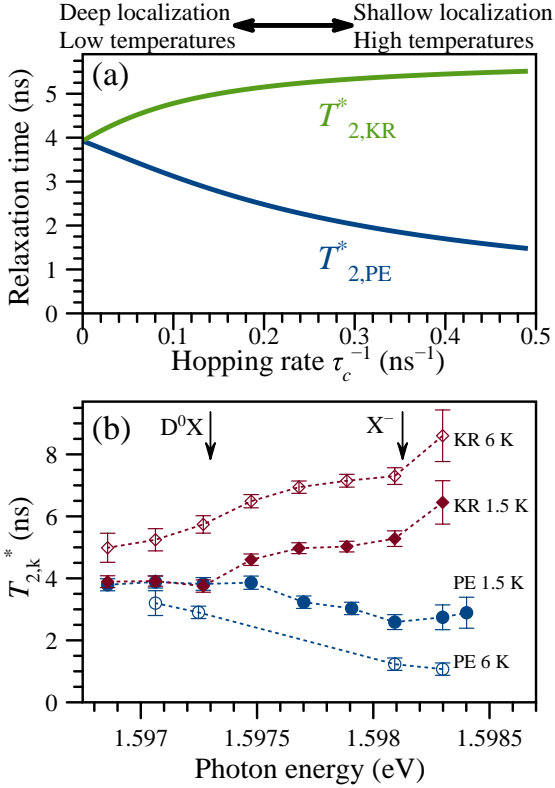}
\caption{(a) Calculated dependence of relaxation times $T_{2,k}^*$ on hopping rate $\tau_c^{-1}$ for $\delta_e=0.1$~ns$^{-1}$, $\Delta g_e = 9\times 10^{-3}$ and $B=0.2$~T. (b) Spectral dependence of relaxation times $T_{2,k}^*$ measured at $T=1.5$ and 6~K for $B=0.2$~T.}
\label{fig4}
\end{figure}

The drastic difference of the PE and KR decay times is related to the specific impacts of the localized electron dynamics on the coherent response, as sketched in Fig.~\ref{fig3}. Hopping of electrons between localization sites destroys the spin grating in Eq.~\eqref{grating}: Once an electron leaves its initial site and arrives at a site with different location $\mathbf{r}$ and frequency $\omega_0$, it no longer contributes to the PE signal. However, it continues to contribute to the KR signal provided that spin is conserved during the course of hopping, since hopping does not change the macroscopic spin polarization and the variation of energy $\Delta E_{\rm hop} \sim k_B T$ is small compared to the spectral width $\Delta \omega_0$. We emphasize that the electron displacement by hopping is small and therefore the spectral diffusion plays decisive role in the decay of PE signal. In our QW the electron spin coherence is controlled by the hyperfine coupling with nuclear spins~\cite{Merkulov-02,glazov:book}. If the hopping processes, characterized by the electron spin correlation time at the localization site $\tau_c$, are efficient such that the electron spin precession rate in the field of nuclear fluctuations $\sqrt{\langle \Omega_N^2\rangle}$ is small, $\sqrt{\langle \Omega_N^2\rangle}\tau_c \lesssim 1$, macroscopic spin relaxation in the ensemble takes place after several hops only with a rate given by $T_{2, \rm KR}^{-1} \sim \langle \Omega_N^2\rangle\tau_c$~\cite{Kavokin-08,glazov:hopping} due to the effect of motional narrowing. By contrast, the PE decays with the time constant $\sim \tau_c$ independent of the spin relaxation mechanism, resulting in $T_{2,\rm PE} \sim \tau_c \ll T_{2,\rm KR}$, in agreement with our findings. The microscopic theory based on the kinetic equation for the spin distribution function in the Appendix B gives the following expressions for the decay times
\begin{equation}
\label{t2:res}
T_{2, \rm KR} = \frac{2\exp{\left[-1/(\delta_e\tau_c)^2\right]}}{\sqrt{\pi}\delta_e \erfc[1/(\delta_e\tau_c)]}, \quad  T_{2,\rm PE} =T_{2,\rm KR} - \frac{2}{\delta_e^2\tau_c},
\end{equation}
with $\delta_e^2 = 2\langle \Omega_N^2\rangle/3$ and $\erfc(x)$ being the error function.
Results of numerical calculations are shown in Fig.~\ref{fig4}(a), where the relaxation times $T_{2,k}^*$ are plotted as a function of the hopping rate, $\tau_c^{-1}$.

The strength of localization governs the hopping rate of the resident electrons. Therefore, it is possible to address different hopping regimes by optical excitation of D$^0$X and X$^-$, i.e. electrons bound to donors (stronger localization) or electrons localized on potential fluctuations (weaker localization), respectively. Furthermore, an increase of temperature will lead to an increase of hopping rate. Consequently, the difference between the relaxation times $T_{2,\rm KR}$ and $T_{2,\rm PE}$ should become even more pronounced. In order to test this conjecture, we have studied the spectral and temperature dependences of $T^*_{2,\rm KR}$ and $T^*_{2,\rm PE}$ at $B=0.2$~T, Fig.~\ref{fig4}(b).

First, we observe that for resonant excitation of D$^0$X (1.5972~eV) at $T=1.5$~K the relaxation times $T^*_{2,\rm KR}$ and $T^*_{2,\rm PE}$ are identical with a value of 4~ns. This is due to the stronger localization of donor bound electrons and therefore the regime of $\langle \Omega_N^2 \rangle \tau_c^2 \gg 1$ is realized in this case~\cite{Korenev-02,Merkulov-02}. Here, the electron spin is lost efficiently at a given donor via the hyperfine coupling and, consequently, hopping is unimportant so that $T_{\rm 2,PE}^* = T_{\rm 2,KR}^*$. Excitation with a larger photon energy addresses electrons with weaker localization and consequently  $T^*_{\rm 2,KR}$ increases while $T^*_{\rm 2,PE}$ decreases, in accordance with our predictions. Second, a temperature increase leads to a similar behavior, which also excludes a possible origin of the $\tau_c$ behavior in the exchange interaction between the resident electrons, because it should be largely independent of temperature. Using the theoretical results in Fig.~\ref{fig4}(a) we determine the hopping time of electrons localized on potential fluctuations to be $\tau_c \approx 5$ and 2~ns for $T=1.5$ and 6~K, respectively. For electrons bound to donors we observe hopping only at $T=6$~K with $\tau_c \approx 5$~ns and $\delta_e \approx 0.1$~ns$^{-1}$.

In conclusion, we have demonstrated that spin-dependent photon echoes represent a powerful tool to access directly the local spin and charge dynamics of resident carriers in semiconductors. Our findings allow one to determine the key transport parameter of a localized system, the hopping rate $\tau_c$ by purely optical means thereby bridging the gap between optical and transport spectroscopy. The electron spin correlation time $\tau_c$ directly controls the dynamical nuclear polarization induced by the hyperfine interaction~\cite{OO-book,glazov:book}, opening up prospects to optimize nuclear spin memories by tailoring electron localization. Finally, we observe suppressed hopping of donor bound electrons at low temperature of 1.5~K where the spin relaxation of the ensemble takes place in the fluctuating nuclear fields. This suggests that the decay of spin-dependent photon echo from donor bound excitons can be further extended by several orders of magnitude using spin echoes which is attractive for applications in quantum communication.

We are grateful to O.~S.~~Ken and V.~L.~~Korenev for useful discussions. We acknowledge the financial support by the Deutsche Forschungsgemeinschaft through the International Collaborative Research Centre TRR 160 (Project A3). S.V.P. thanks the Russian Foundation for Basic Research (RFBR) (Project No. 19-52-12046) and St. Petersburg State University (Project No. 11.34.2.2012 and grant ID 40847559). L.E.G. and M.M.G. acknowledge partial support from RFBR (Project No. 19-52-12038); L.E.G. is grateful to the Foundation for advancement of theoretical physics and mathematics ``BASIS''. The research in Poland was partially supported by the Foundation for Polish Science through the IRA Programme co-ﬁnanced by the EU within SG OP and by the National Science Centre through Grants No. 2017/25/B/ST3/02966 and 2018/30/M/ST3/00276.

%\newpage

\appendix

%\tableofcontents

\section{Spectral dependence of $g$ factor}

The following experimental techniques can be used to evaluate the $g$ factor of resident electrons: Time-resolved Kerr rotation (KR), resonance spin amplification (RSA) and 3-pulse spin dependent photon echoes (PEs). The first and third technique are described in detail in the main text. Resonant spin amplification exploits an accumulative phenomenon, which results in enhancement of the pump-probe KR amplitude when the repetition rate of the exciting laser pulses $1/T_R$ corresponds to a multiple of the Larmor precession frequency $\Omega_L /2 \pi$ in case the spin dephasing time $T^*_{\rm 2,KR}$ exceeds the pulse repetition period $T_R$ significantly~\cite{Kikkawa-98}. In our case $T_R = 13.2$~ns and, therefore, the latter condition is fulfilled. Typically, the RSA signal is measured by scanning the magnetic field and thereby tuning $\Omega_L$, and detecting the KR signal at a fixed small negative delay time right before the pump pulse arrival. Thus, while KR and PE transients capture the spin dynamics occurring subsequent to spin initialization, the RSA signal reflects the accumulated signal, which is result of excitation with a periodic sequence of optical pulses. Here, we compare the results of all three methods and their spectral dependences.

\begin{figure}[htp]
.ö\includegraphics[width=0.85\columnwidth]{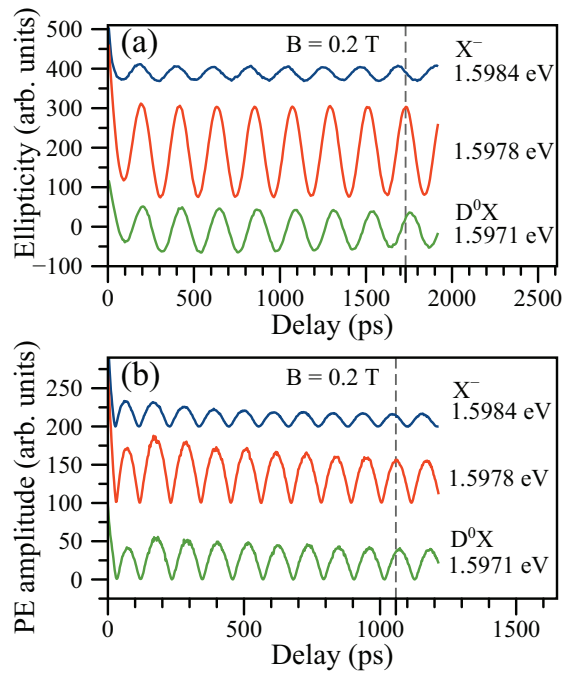}
\caption{Time-resolved KR signals, detected in ellipticity, (a) and 3-pulse spin-dependent PE signals for $\tau_{12}=26.7$~ps (b), in both settings measured for resonant excitation of D$^0$X (1.5971~eV) and of X$^-$ ($1.5984$~eV) as well as for an intermediate photon energy 1.5978~eV. The signals are vertically shifted for clarity. $T=1.5$~K and $B=0.2$~T. }
\label{fig:TRKR+PE}
\end{figure}

\begin{figure}[htp]
\includegraphics[width=0.9\columnwidth]{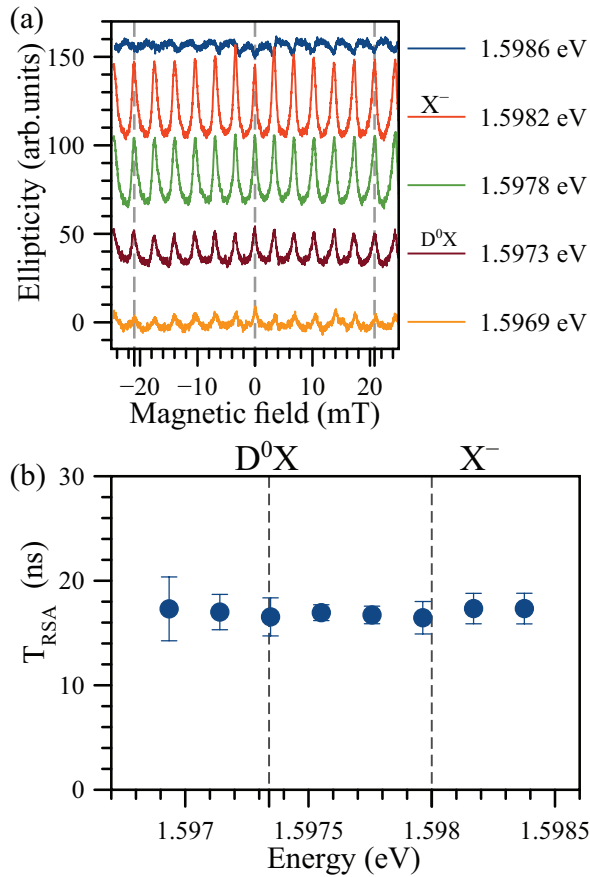}
\caption{(a) RSA signals measured for different photon energies $\hbar\omega$. $T=1.5$~K. The delay time between the pump and probe pulses is set to $t\approx -10$~ps. (b) Spectral dependence of spin relaxation time $T_{\rm RSA}$, extracted from the halfwidth of the RSA peaks $B_{1/2}$.}
\label{fig:RSA}
\end{figure}

Transient KR and PE signals are shown in Figs.~\ref{fig:TRKR+PE}(a) and \ref{fig:TRKR+PE}(b), respectively, recorded in both cases for three different photon excitation energies. The difference between the Larmor precession frequencies for resonant excitation of X$^-$ and D$^0$X complexes is well pronounced. This difference originates from the difference in the $g$ factor of resident electrons localized in potential fluctuations and bound to donors~\cite{Salewski-17}.

Figure~\ref{fig:RSA} shows signals obtained in RSA measurements. The separation between neighboring RSA peaks corresponds to $\Delta B = 2\pi \hbar / g_e \mu_B T_R$,  from which the $g$ factor of resident electrons can be as well extracted. Additionally, the halfwidth of the peaks at half maximum $B_{1/2} = \hbar / g_e \mu_B T_{\rm RSA}$ allows us to determine the transverse spin relaxation time $T_{\rm  RSA}$. Interestingly, according to Fig.~\ref{fig:RSA}(a), the position of all peaks and consequently the $g$ factor of the resident electrons are independent of the excitation photon energy $\hbar\omega$. Also, the spin relaxation time shows no spectral dependence and is constant at a value of about 20~ns, as shown in Fig.~\ref{fig:RSA}(b). This is in contrast to time-resolved measurements of KR and PE, see Fig.~4(b) in the main text.

\begin{figure}[htp]
\includegraphics[width=0.9\columnwidth]{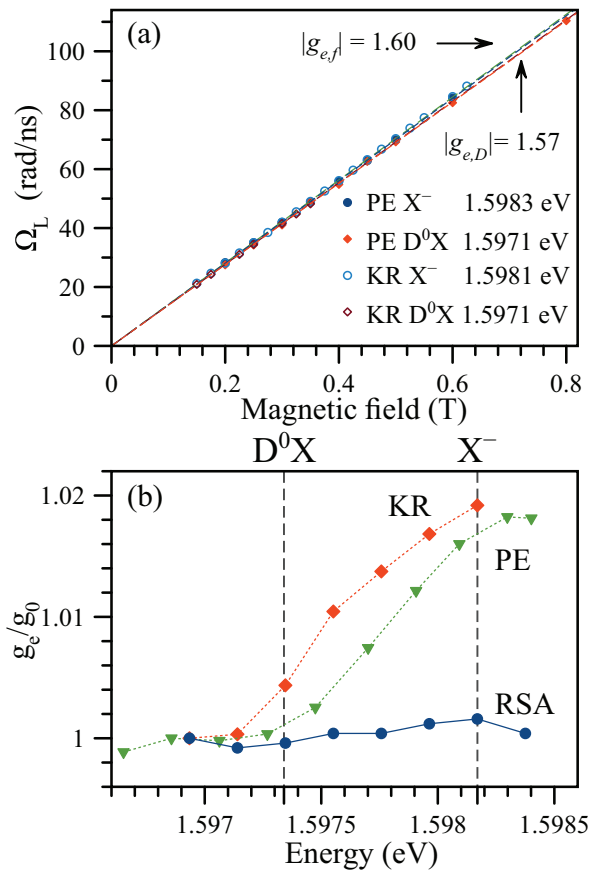}
\caption{(a) Magnetic field dependence of Larmor precession frequency $\Omega_L$ evaluated from KR (open symbols) and PE (solid symbols) transients under resonant excitation of the X$^-$(circles) and D$^0$X (squares) complexes. (b) Spectral dependence of normalized $g$ factor $g_e/g_0$ extracted using three different methods: red squares – KR, green triangles - PE and blue circles - RSA. $T=2$~K. PE and KR data are measured at $B=0.2$~T.}
\label{fig:SpcDep}
\end{figure}

The results of the $g$ factor measurements are summarized in Fig.~\ref{fig:SpcDep}. The absolute value of the $g$ factor can be determined with high precision from a linear fit to the magnetic field dependence of the Larmor precession frequency. This allows us to minimize the error related to the uncertainty in setting the value of the external magnetic field. The results evaluated from  KR and PE transients for resonant excitation of D$^0$X and X$^-$ are shown in Fig.~\ref{fig:SpcDep}(a). Taking into account that the electron $g$ factor is negative (see, e.g., Ref.~\cite{Sirenko-97}) we obtain $g_{e,f} = -1.60$ and $g_{e,D} = -1.57$ for the resident electrons localized on potential fluctuations and bound to donors, respectively. The accuracy of this evaluation is better than 0.5\%.
% $g$ factors, extracted from the fit has a value of 1.595 (PE X−), 1.571 (PE D0X), 1.602 (KR X−) and 1.575 (KR D0X).
We do not compare here the absolute values of $g_e$ with the RSA data because the RSA scans were performed at low magnetic fields where the measurement accuracy of the absolute value of the magnetic field is quite low resulting in a too large error (about 5\%).

The dependence of the $g$ factor on photon energy $\hbar\omega$ can be extracted also with high precision for a fixed external magnetic field strength. In this case we are sensitive to relative changes of the $g$ factor and therefore the results are normalized by $g_0$ which corresponds to the magnitude of $g_e$ measured by each of the methods at 1.597~eV. These results are shown in Fig.~\ref{fig:SpcDep}(b). The spectral dependencies from KR and PE are very similar. It follows that the $g$ factor increases by about 2\% when the photon energy is tuned from the D$^0$X to the X$^-$ resonance. This is in accordance with the $g$ factor values evaluated from the magnetic field dependencies in Fig.~\ref{fig:SpcDep}(a). In contrast, as mentioned RSA does not reveal any spectral dependence of the $g$ factor. Such a behavior can be related to different spin dephasing times for electrons bound to donors and electrons localized on potential fluctuations. The RSA signal from the resident electrons with longest time is the most pronounced. In addition, spin transport between the two different types of localization sites can cause a flattening of the $g$ factor spectral dependence in the RSA measurements. The transport can occur due to hopping of carriers between the localization sites, due to diffusion, or due to exchange interaction between carriers.

\section{Theory}

\subsection{Formation of 3-pulse spin-dependent\\ photon echo signal}

We consider an ensemble of resident electrons localized on random fluctuations of the quantum well potential caused by interface imperfections, as well as by the presence of donors both in the barriers and in the well. Let $\varepsilon$ be the electron energy and $\mathcal D(\varepsilon)$ be the density of (localized) states related to disorder. We assume that the incident photons result in the formation of three particle complexes consisting of two electrons and a hole, i.e. of negatively charged trions or donor-bound excitons. Due to the potential fluctuations the energy of the complexes $\varepsilon_3$ also fluctuates. As a result, the energy of the optical transition defined by the difference of the three-particle complex energy and the electron energy
\begin{equation}
\label{energy}
\hbar\omega_0 = \varepsilon_3 - \varepsilon,
\end{equation}
fluctuates as well. Therefore, the incident optical pulses with a certain spectral width excite a broad distribution of localized states.
Thus, after initialization with two cross-linearly polarized laser pulses (e.g. in a HV sequence where H and V correspond to horizontal and vertical polarization, respectively, as shown in Fig.~1(b) of the main text), the spin density is energy- and coordinate-dependent, so that a spin grating is formed~\cite{Langer-12} [cf. Eq.~(1) of the main text]:
\begin{multline}
\label{FMW:init}
S_y^0( \varepsilon, \mathbf r) {-\mathrm i S_z^0( \varepsilon, \mathbf r)} \\ \propto
	{\mathrm i \exp\left({\mathrm i\Omega_L\tau_{12}\over 2}\right)
	%\cos{(\mathbf{k}_{\rm FWM} \cdot \mathbf r)}
	\cos{\left(\omega_0\tau_{12} + \mathbf k_{2,\parallel}\mathbf r - \mathbf k_{1,\parallel} \mathbf r \right)}} f(\omega_0).
\end{multline}
Here the $S^0_{\alpha}( \varepsilon, \mathbf r)$ ($\alpha=y,z$) are the initial distributions of the electron spin components, and $f(\omega_0)$ is the function which determines the spectral distribution of the excited spins (this function is omitted for brevity in Eq.~(1) of the main text): It is given by the spectral function of the laser pulse if in comparison the inhomogeneous broadening is larger, or by the inhomogeneity of the system if the inhomogeneous broadening is smaller than the spectral width of the pulses. Note that in Eq.~\eqref{FMW:init} the resonance frequency $\hbar\omega_0$ depends on the electron localization energy $\varepsilon$ via Eq.~\eqref{energy}, and the electron Larmor precession frequency $\Omega_L$ also depends on $\varepsilon$ due to the energy dependence of the $g$ factor~\cite{ivchenko05a}. We recall that $\tau_{12}$ is the time delay between the first two pulses, $\mathbf k_{1,\parallel}$ and $\mathbf k_{2,\parallel}$ are the in-plane components of the wavevectors of the first and the second pulse. In the derivation of Eq.~\eqref{FMW:init} we disregarded the nuclear field fluctuations which are characterized by the Larmor precession frequency $\Omega_N$, assuming that
\begin{equation}
\label{cond:B}
\Omega_L \gg \sqrt{\langle\Omega_N^2\rangle}.
\end{equation}

Our system is strongly inhomogeneous, therefore no average macroscopic spin polarization is formed.
After the time delay $\tau_{23}$, the third V-polarized pulse arrives with in-plane wavevector $\mathbf k_{2,\parallel}$, creating interband coherence which rephases exactly at the time delay $\tau_{12}$ after the third pulse. The photon echo signal is detected along the direction $2\mathbf k_{2,\parallel} - \mathbf k_{1,\parallel}$. For a sufficiently wide distribution of resonance frequencies the induced polarization can be recast as:
\begin{multline}
\label{HVVH}
	P_\text{HVVH}(\tau_{23}) \propto \int d\varepsilon \int d\mathbf r \left[S_y(\varepsilon,\mathbf r; \tau_{23}) - \mathrm i S_z(\varepsilon,\mathbf r; \tau_{23})\right]   \\ \times
	\left[S_y^0(\varepsilon, \mathbf r)+ \mathrm i S_z^0(\varepsilon, \mathbf r)\right].
\end{multline}
It follows from Eqs.~\eqref{FMW:init} and \eqref{HVVH} that the temporal evolution of the photon echo signal is determined by the evolution of the initial distribution~\eqref{FMW:init}. This evolution can be related to the following effects:
\begin{enumerate}
\item Spin dephasing in the ensemble through the inhomogeneous spread of the electron $g$ factor and the corresponding spread of $\Omega_L$.
\item Spin dynamics at a given localization site, i.e., spin precession in the field of the nuclear fluctuations or spin flip due to the electron-phonon interaction.
\item Electron motion in real space, i.e., electron diffusion resulting in decay of the $\cos{\left(\omega_0\tau_{12} + \mathbf k_{2,\parallel}\mathbf r - \mathbf k_{1,\parallel} \mathbf r \right)}$ factor in Eq.~\eqref{FMW:init} due to its coordinate dependence.
\item Electron diffusion in energy space resulting in variation of $\varepsilon$ and decay of the factor $\cos{\left(\omega_0\tau_{12} + \mathbf k_{2,\parallel}\mathbf r - \mathbf k_{1,\parallel} \mathbf r \right)}$ in Eq.~\eqref{FMW:init} due to the $\omega_0(\varepsilon)$ dependence in Eq.~\eqref{HVVH} and the corresponding dependencies of the electron $g$ factor and Larmor frequency $\Omega_L$ on $\varepsilon$ as well; exchange-interaction induced two-electron flip-flop processes~\cite{Kavokin-08} also result in temporal variation of Eq.~\eqref{FMW:init}.
\item Additionally, electron spin precession during the course of hopping between localization sites~\cite{rotation1, rotation2} also results in the evolution of $\mathbf S(\varepsilon, \mathbf r;t)$.
\end{enumerate}
Thus, any process resulting in decay of the spin grating results in a decay of the PE signal. Below we provide a microscopic model of the decay of the spin grating due to electron hopping processes in the presence of hyperfine interaction with nuclear spin fluctuations.

\subsection{Dynamics of the spin grating}

The kinetic equation for the spin density $\mathbf S \equiv \mathbf S( \varepsilon, \mathbf r; t)$ reads
\begin{equation}
\label{kin:tot}
\frac{\partial \mathbf S}{\partial t} + \mathbf S\times [\mathbf \Omega_N(\varepsilon,\mathbf r) + \mathbf \Omega_L(\varepsilon)]  = \mathbf Q\{ \mathbf S\}.
\end{equation}

Here we characterize the electron localization site by two parameters, its coordinate $\mathbf r$ and its site energy $\varepsilon$. Accordingly, the electron spin precession frequency is determined by two contributions: $ \mathbf \Omega_N(\varepsilon,\mathbf r)$ due to the nuclear spin fluctuations and $\mathbf \Omega_L(\varepsilon) = g_{e}(\varepsilon) \mu_B \mathbf B/\hbar$ due to the Zeeman splitting in the external magnetic field, where $g_{e}(\varepsilon)$ is the electron Land\'e factor, which is assumed to depend only on the electron localization energy. The collision integral $\mathbf Q\{ \mathbf S\}$ describes the electron jumps both in real and energy space,
\begin{multline}
\label{collision}
\mathbf Q\{ \mathbf S\} = \int d\mathbf r' d\varepsilon \left[w(\varepsilon,\mathbf r;\varepsilon',\mathbf r') \mathbf S(\varepsilon',\mathbf r';t)\right. \\
\left.- w(\varepsilon',\mathbf r';\varepsilon,\mathbf r) \mathbf S(\varepsilon,\mathbf r; t)\right],
\end{multline}
with $w(\varepsilon',\mathbf r',\varepsilon,\mathbf r)$ being the probability per unit time of electron hopping from the site $\varepsilon,\mathbf r$ to the site $\varepsilon',\mathbf r'$,
\[
\frac{w(\varepsilon',\mathbf r';\varepsilon,\mathbf r)}{w(\varepsilon,\mathbf r;\varepsilon',\mathbf r')} = \exp{\left(\frac{{\varepsilon - \varepsilon'}}{k_B T} \right)},
\]
where $T$ is the lattice temperature and $k_B$ is the Boltzmann constant.

The spin relaxation at a given site via, e.g., electron-phonon interaction can be taken into account by including an additional term $- \mathbf S/\tau_s$ on the right-hand side of Eq.~\eqref{kin:tot}. Estimates show that the electron-phonon interaction manifests itself only at much higher magnetic fields than applied here and therefore can be neglected~\cite{glazov:book}. In derivation of Eq.~\eqref{kin:tot} we disregarded the electron spin precession during hopping between  localization sites and the exchange interaction between the electrons. Thus, we consider the hyperfine interaction as the sole source of electron spin relaxation in our system.

\begin{figure}[htp]
\includegraphics[width=\columnwidth]{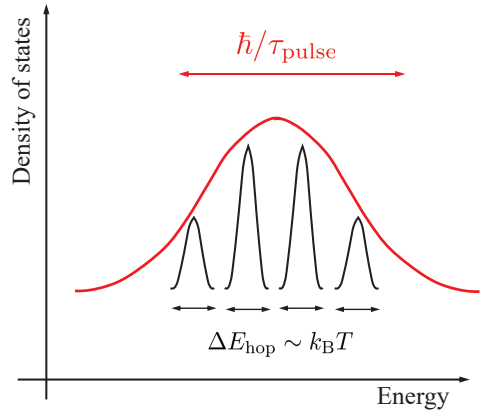}
\caption{Sketch of the hopping processes in the energy space: The pump pulse excites rather a wide distribution of electrons with spectral width $\sim \hbar/\tau_p \sim 1$~meV, where $\tau_{\rm pulse}$ is the pulse duration. Hopping occurs within much narrower energy ranges $\Delta E_{\rm hop} \sim k_B T \sim 0.15$~meV at $T =1.5$~K.}
\label{fig:hopping}
\end{figure}

As a rule, in systems with localized electrons a wide distribution of hopping probabilities exists~\cite{ES:book}, leading to nontrivial spin dynamics~\cite{raikh, shumilin}. The description of these effects goes far beyond the scope of the present paper and we use instead a simplified form of the collision integral characterized by a single correlation time $\tau_c$, which allows us to obtain an analytical solution for the spin dynamics~\cite{glazov:hopping,glazov:book}:
\begin{equation}
\label{diff}
\bm Q\{ \bm S\} = \frac{1}{\tau_c} \left[\int_{\varepsilon'\in \Delta E_{\rm hop}} d\varepsilon' \mathcal D(\varepsilon') \bm S(\varepsilon') - \bm S(\varepsilon)\right].
\end{equation}
Here the integration is carried out over the stripe of energies with width $\Delta E_{\rm hop}\sim k_B T$ involved in the hopping process, $\int_{\varepsilon'\in \Delta E_{\rm hop}} d\varepsilon' \mathcal D(\varepsilon') =1$, see Fig.~\ref{fig:hopping}. Note that the coordinate dependence of the spin distribution function $\bm S$ can be omitted: Our estimates show that the product $\mathbf k_{\rm FWM} \Delta \mathbf r$, where $\Delta \mathbf r$ is the electron displacement by one hopping event, is negligibly small. Thus, only energy diffusion and hyperfine interaction are important for decay of the photon echo signal.

We also assume that the condition~\eqref{cond:B} is fulfilled, thus we neglect the nuclear spin fluctuations transverse to the external magnetic field. Hence, the pseudovector kinetic equation~\eqref{kin:tot} can be reduced to a scalar equation for the function
\[
S_+(\varepsilon,\mathbf r) = S_y(\varepsilon,\mathbf r) + \mathrm i S_z(\varepsilon,\mathbf r),
\]
namely:
\begin{multline}
\label{tc:simplified}
\frac{\partial S_+(\varepsilon)}{\partial t} + \mathrm i (\Omega_L + \Omega_{N,x}) S_+ (\varepsilon) =\\
   \frac{1}{\tau_c} \left[\int d\varepsilon' \mathcal D(\varepsilon') S_+(\varepsilon') -  S_+(\varepsilon)\right] + \delta(t) G_+(\varepsilon).
\end{multline}
Here $G_+(\varepsilon) \propto S_y^0 + \mathrm i S_z^0$, cf. Eq.~\eqref{FMW:init}, is proportional to the spin distribution generated by the first two pulses, and we take into account that $\tau_{23} \gg \tau_{12}$ which allows us to assume that the initial spin distribution is formed instantaneously compared to the time scale of its further evolution.

We present the solution of Eq.~\eqref{tc:simplified} under the additional simplification that the variation of the electron $g$ factor within the hopping energy range $\Delta \varepsilon$ is negligible:
\begin{equation}
\label{cond:g}
\left |\frac{dg_{e}}{d\varepsilon} \Delta \varepsilon \mu_B B\right| \ll  \sqrt{\langle\Omega_N^2\rangle}.
\end{equation}
Thus, we neglect the variation of the electron spin precession frequencies in the external magnetic field during hopping and take into account the $g$ factor spread, while calculating the photon echo decay by averaging over the Larmor frequency distribution. This assumption is confirmed according to the spectral dependence of the $g$ factor obtained from experiment (Fig.~\ref{fig:SpcDep}). Introducing the notation
\begin{equation}
\label{ST}
S_t = \int d\varepsilon' \mathcal D(\varepsilon') S_+(\varepsilon'),
\end{equation}
and making use of the Fourier transform over time we arrive at the algebraic equation
\begin{equation}
\label{tc:simplified:1}
\left[-\mathrm i (\Omega -\Omega_L) + \tau_c^{-1} + \mathrm i \Omega_{N,x}(\varepsilon) \right]S_+ =  \tau_c^{-1} S_t +  G_+(\varepsilon),
\end{equation}
which is readily solved with the result
\begin{equation}
\label{tc:simplified:2}
S_+(\Omega) =  \frac{\tau_c^{-1} S_t +  G_+(\varepsilon)}{-\mathrm i  (\Omega -\Omega_L)  + \tau_c^{-1} + \mathrm i\Omega_{N,x}(\varepsilon)}.
\end{equation}
Solving the self-consistency equation~\eqref{ST} and averaging over random realizations of the nuclear fields we obtain
\begin{subequations}
\label{tc:simplified:3}
\begin{align}
& S_t =  \frac{\tau_c}{1-\mathcal A_{\Omega}^+}\left\langle\int\frac{ d\varepsilon  \mathcal D(\varepsilon)  G_+(\varepsilon)}{1-\mathrm i (\Omega-\Omega_L)\tau_c  + \mathrm i \Omega_{N,x}(\varepsilon)\tau_c}\right\rangle, \\
& \mathcal A_{\Omega}^+ = \left\langle \int \frac{d\varepsilon \mathcal D(\varepsilon)}{1-\mathrm i (\Omega-\Omega_L)\tau_c  + \mathrm i \Omega_{N,x}(\varepsilon)\tau_c}\right\rangle.
\end{align}
\end{subequations}
Here $\langle K(\Omega_N) \rangle = \int d\mathbf \Omega_{N} K(\Omega_N) \mathcal F(\mathbf \Omega_N)$, where $\mathcal F(\mathbf \Omega_N)$ is the distribution function of the nuclear spin fluctuations, which are static on the time scale of the electron spin dynamics~\cite{glazov:book}.
Using the explicit form of the distribution for a given component of the nuclear field $\mathcal F_x(\Omega_{N,x}) = \exp{[-\Omega^2_{{N,x}}/\delta_e^2]}/(\sqrt{\pi} \delta_e)$ with $\delta_e^2 = 2\langle\Omega_N^2\rangle/3$~\cite{glazov:book} we obtain:
\begin{equation}
\label{a+}
\mathcal A_{\Omega}^+ = \frac{\sqrt{\pi}}{\tau_c\delta_e} \exp{\left(\frac{1}{\delta_e^2\tau_{\Omega-\Omega_L}^2}\right)}\erfc{\left(\frac{1}{\delta_e\tau_{\Omega-\Omega_L}}\right)}.
\end{equation}
Here $\tau_{\Omega}=\tau_c/(1-\mathrm i\Omega\tau_c)$.
We also present the asymptotic expressions for the function $\mathcal A_{\Omega}^+$:
\[
\mathcal A_{\Omega}^+ = \begin{cases}
[1 - \mathrm i (\Omega - \Omega_L)\tau_c + {\delta_e^2\tau_c^2}/{2}]^{-1}, \quad \delta_e \tau_c \ll 1,\\
\frac{\sqrt{\pi}}{\tau_c\delta_e}e^{-(\Omega-\Omega_L)^2/\delta_e^2}\erfc{[-\mathrm i(\Omega-\Omega_L)/\delta_e]}, \quad \delta_e \tau_c \gg 1.
\end{cases}
\]

According to Eq.~\eqref{HVVH} the photon echo signal is given by the convolution of $S_y(\varepsilon, \mathbf r; \tau_{23}) -\mathrm i S_z(\varepsilon, \mathbf r; \tau_{23})$ and the function $G_+$ averaged over the nuclear fluctuations. Since the energy-averaged value of $G_+$ is equal to zero, only the second term in the numerator of Eq.~\eqref{tc:simplified:2} contributes. As a result
\begin{equation}
\label{phvvh:calc}
P_\text{HVVH}(\tau_{23}) \propto \int_{-\infty}^\infty \frac{d\Omega}{2\pi} \Re{\left( \mathcal A_{\Omega}^+ e^{-\mathrm i \Omega\tau_{23} + \mathrm i \phi}\right)},
\end{equation}
where $\phi$ is the phase related to the spin initialization conditions.
The spread of the electron $g$ factors can be accounted for by averaging Eq.~\eqref{phvvh:calc} over the distribution of $\Omega_L$.

The function $\mathcal A_{\Omega}^+$ has a sharp peak at $\Omega \approx \Omega_L$ so that in the vicinity of $\Omega \approx \Omega_L$ it can be approximated by
\begin{equation}
\label{pole}
\mathcal A_{\Omega}^+ \approx \frac{A_{\Omega_L}}{1-\mathrm i (\Omega-\Omega_L)T_{\rm 2,PE}},
\end{equation}
thus $P_\text{HVVH}(\tau_{23})$ oscillates with the delay $\tau_{23}$ as $\cos{(\Omega\tau_{23} -\phi)}\exp{(-\tau_{23}/T_{\rm 2,PE})}$. After some algebra we obtain
\begin{equation}
\label{t2pe}
T_{\rm 2,PE} =\frac{2}{\delta_e}\left(\frac{\exp{\left[-1/(\delta_e\tau_c)^2\right]}}{\sqrt{\pi} \erfc[1/(\delta_e\tau_c)]} -\frac{1}{\delta_e\tau_c}\right),
\end{equation}
see Eq.~(2) of the main text.
In the relevant limiting cases we obtain
\begin{equation}
\label{t2pe:a}
\frac{1}{T_{\rm 2,PE}} \approx \begin{cases}
{\sqrt{\pi} \delta_e}/{2}, \quad \delta_e \tau_c \gg 1,\\
1/\tau_c, \quad \delta_e \tau_c \ll 1,
\end{cases}
\end{equation}
Indeed, if $\delta_e \tau_c \gg 1$ the spin is lost already at a given localization site due to the nuclei-induced spin precession with rate $\sim \delta_e$ and hopping is unimportant. We note that in this regime the spin decoherence is described by a Gaussian law and the above expressions provide a convenient analytical interpolation. In the motional narrowing regime, $\delta_e \tau_c \ll 1$, the photon echo decays with the rate $\sim 1/\tau_c$, in agreement with the qualitative analysis.

\subsection{Kerr rotation dynamics}

The model developed above can be also used to describe the time-resolved Kerr rotation signal. In this case a macroscopic, i.e. on average non-zero, spin polarization is generated. The dynamics of the Kerr rotation signal is given by the Fourier transform of the total spin $S_t$ in Eq.~\eqref{tc:simplified:3}. Thus, instead of Eq.~\eqref{phvvh:calc}
\begin{equation}
\label{KR:calc}
S_\text{KR}(t) \propto \int_{-\infty}^\infty \frac{d\Omega}{2\pi}  \Re{\left(\frac{\mathcal A_{\Omega}^+}{1-\mathcal A_{\Omega}^+}e^{-\mathrm i \Omega t}\right)}.
\end{equation}
As a result, in order to find the decay of the Kerr rotation signal we have to analyze the poles of $\mathcal A_{\Omega}^+/({1-\mathcal A_{\Omega}^+})$ at $\Omega \approx \Omega_L$. Along the same lines as above we arrive at the following expression for $T_{\rm 2,KR}$ [cf. Eq. (2) of the main text]:
\begin{equation}
\label{t2kr}
T_{\rm 2,KR} =\frac{2\exp{\left[-1/(\delta_e\tau_c)^2\right]}}{\sqrt{\pi}\delta_e \erfc[1/(\delta_e\tau_c)]},
\end{equation}
with the following asymptotes
\begin{equation}
\label{t2kr:a}
\frac{1}{T_{\rm 2,KR}} \approx \begin{cases}
{\sqrt{\pi} \delta_e}/{2}, \quad \delta_e \tau_c \gg 1,\\
{\delta_e^2\tau_c}/2, \quad \delta_e \tau_c \ll 1.
\end{cases}
\end{equation}
Naturally, for long correlation times, $\tau_c \delta_e \gg 1$, both Kerr rotation and photon echo signals decay with the same rate because the electron spin is lost before the electron hops. By contrast, for short correlation times, $\tau_c \delta_e \ll 1$, the motional narrowing results in a significant slow-down of the Kerr rotation decay as compared with the decay of the photon echo.

Note that in order to  include the spread of the electron $g$ factor values it is sufficient to average Eqs.~\eqref{phvvh:calc} and \eqref{KR:calc} over the distribution of $g_e$, under the condition~\eqref{cond:g}. Roughly this corresponds to the replacement
\[
\frac{1}{T_{\rm 2,PE/KR}}\to \frac{1}{T_{\rm 2,PE/KR}^*} = \frac{1}{T_{\rm 2,PE/KR}} + \frac{\Delta g_e \mu_B B}{\hbar},
\]
where $\Delta g_e$ is the characteristic spread of the $g$ factor values. We also note that for strong electron-electron exchange interaction the correlation time $\tau_c$ can be controlled by the exchange interaction between neighboring occupied localization sites. The correlation rate can be estimated in this regime as $1/\tau_c \sim J/\hbar$, where $J$ is the exchange constant, and does not depend on temperature~\cite{Kavokin-08}. In this regime the above expressions hold, however, the strong temperature dependence of $T_{\rm 2,PE/KR}^*$ observed in the experiments, see Fig. 4 of the main text, rules out the exchange interaction as origin of the observations.
Also, if phonon-induced spin relaxation at a given site were important, it would affect $T_{\rm 2,PE}$ and $T_{\rm 2,KR}$ in the same way.

\subsection{Details of numerical calculation}

Although Eqs.~\eqref{t2pe} and~\eqref{t2kr} provide analytical expressions for the decay times of the signals, they become, strictly speaking, invalid at $\tau_c \to \infty$ due to the Gaussian decay of the signals. In this case, the pole approximation~\eqref{pole} [and analogously the one for $\mathcal A_{\Omega}^+/({1-\mathcal A_{\Omega}^+})$ in the case of Kerr rotation] becomes inapplicable. Mathematically, this is because the function $\exp{[-(\Omega - \Omega_L)^2]}$ does not have poles in the complex plane of $\Omega$. Thus, in order to determine the decay time of the signals we calculated $1/T_{\rm 2,PE}$ and $1/T_{\rm 2,KR}$ by finding numerically the half-width at half maximum of $\mathcal A_{\Omega}^+$ (for the photon echo) and of $\mathcal A_{\Omega}^+/({1-\mathcal A_{\Omega}^+})$ (for the Kerr rotation). In the motional narrowing regime $\delta_e\tau_c \ll 1$ we recover the same asymptotes as in Eq.~\eqref{t2pe:a} and \eqref{t2kr:a}. For $\delta_e \tau_c \gg 1$ we have the same decay rates for the photon echo and Kerr rotation, $\delta_e$ (within $12\%$ of the analytical result). In the numerical calculations shown in {Fig.~4} of the main text we have also included the $g$ factor spread.


\begin{thebibliography} {99}

\bibitem{Dyakonov-book} {\it Spin Physics in Semiconductors}, edited by M. I. Dyakonov (Springer, Berlin, 2011).

\bibitem{Zutic-04} I. Zutic, J. Fabian and S. Das Sarma, {\it Spintronics: Fundamentals
and applications}, Rev. Mod. Phys. {\bf 76}, 323 (2004).

\bibitem{Awschalom-09} J. D. Koralek, C. P. Weber, J. Orenstein, B. A. Bernevig, Shou-Cheng Zhang, S. Mack, and D. D. Awschalom {\it Emergence of the persistent spin helix in semiconductor quantum wells}, Nature {\bf 458}, 610 (2009).

\bibitem{Kukushkin-16} A.~S. Zhuravlev, V.~A. Kuznetsov, L.~V. Kulik, V.~E. Bisti, V.~E. Kirpichev, I.~V. Kukushkin, and S. Schmult, {\it Artificially constructed plasmarons and plasmon-exciton molecules in 2D metals}, Phys. Rev. Lett. {\bf 117}, 196802 (2016).

\bibitem{Belykh-18} V. V.~Belykh, A. Yu.~Kuntsevich, M. M.~Glazov, K. V.~Kavokin, D. R.~Yakovlev, and M. Bayer,{\it Quantum interference controls the electron spin dynamics in n-GaAs}, Phys. Rev. X {\bf 8}, 031021 (2018).

\bibitem{glazov:book} M.~M. Glazov, \emph{Electron \& Nuclear Spin Dynamics in Semiconductor Nanostructures} (Oxford University Press, Oxford, 2018).

\bibitem{Kavokin-02} R. I. Dzhioev, K. V. Kavokin, V. L. Korenev, M. V. Lazarev, B. Ya. Meltser, M. N. Stepanova, B. P. Zakharchenya, D. Gammon, and D. S. Katzer, {\it Low-temperature spin relaxation in n-type GaAs}, Phys. Rev. B {\bf 66}, 245204 (2002).

\bibitem{Korenev-02}
R.~I.~Dzhioev, V.~L.~Korenev, I.~A.~Merkulov, B.~P.~Zakharchenya, D.~Gammon, Al.~L.~Efros, and D.~S.~Katzer, {\it Manipulation of the spin memory of electrons in n-GaAs}, Phys. Rev. Lett.  {\bf 88},  256801 (2002).

\bibitem{Cundiff-06}
S. G. Carter, Z. Chen, and S. T. Cundiff, {\it Optical measurement and control of spin diffusion in n-doped GaAs quantum wells}, Phys. Rev. Lett. {\bf 97}, 136602 (2006).

\bibitem{Golub-17} D.~S. Smirnov and L.~E. Golub, {\it Electrical spin orientation, spin-galvanic, and spin-Hall effects in disordered two-dimensional systems}, Phys. Rev. Lett. {\bf 118}, 116801 (2017).

\bibitem{Golub-18} A. V. Shumilin, D. S. Smirnov, and L. E. Golub, {\it Spin-related phenomena in the two-dimensional hopping regime in magnetic field}, Phys. Rev. B {\bf 98}, 155304 (2018).

\bibitem{Yu-15} Z.G. Yu, {\it Spin Hall effect in disordered organic solids}, Phys. Rev. Lett. {\bf 115}, 026601 (2015).

\bibitem{Awschalom-book}  {\it Semiconductor Spintronics and Quantum Computation } edited by D. D. Awschalom, D. Loss, and N. Samarth, (Springer, Berlin, 2002).

\bibitem{OO-book} {\it Optical Orientation}, edited by F. Meier and B. Zakharchenya
(North-Holland, New York, 1984).

\bibitem{Amand-02} S. Cortez, O. Krebs, S. Laurent, M. Senes, X. Marie, P. Voisin, R. Ferreira, G. Bastard, J-M. G\'erard, and T. Amand, {\it Optically driven spin memory in n-doped InAs-GaAs quantum dots}, Phys. Rev. Lett. {\bf 89}, 207401 (2002).

\bibitem{Crooker-97} S.~A.~Crooker, D.~D.~Awschalom, J.~J.~Baumberg, F.~Flack, and N.~Samarth, {\it Optical spin resonance and transverse spin relaxation in magnetic semiconductor quantum wells}, Phys. Rev. B {\bf 56}, 7574 (1997).

\bibitem{Yugova2009} I.~A.~Yugova, M.~M.~Glazov, E.~L.~Ivchenko, and Al.~L.~Efros, {\it Pump-probe Faraday rotation and ellipticity in an ensemble of singly charged quantum dots}, Phys. Rev. B {\bf 80}, 104436 (2009).

\bibitem{Glazov2010} M.~M~Glazov, I.~A.~Yugova, S.~Spatzek, A.~Schwan, S.~Varwig, D.~R.~Yakovlev, D.~Reuter, A.~D.~Wiek, and M.~Bayer, {\it Effect of pump-probe detuning on the Faraday rotation and ellipticity signals of mode-locked spins in (In,Ga)As/GaAs quantum dots}, Phys. Rev. B {\bf 82}, 155325 (2010).

\bibitem{Langer-12} L.~Langer, S.~V.~Poltavtsev, I.~A.~Yugova, D.~R.~Yakovlev, G.~Karczewski, T.~Wojtowicz, J.~Kossut, I.~A.~Akimov, and M.~Bayer, {\it Magnetic-field control of photon echo from the electron-trion system in a CdTe quantum well: Shuffling coherence between optically accessible and inaccessible states}, Phys. Rev. Lett. {\bf 109}, 157403 (2012).

\bibitem{Langer-14} L.~Langer, S.V.~Poltavtsev, I.A.~Yugova, M.~Salewski, D.R.~Yakovlev, G.~Karczewski, T.~Wojtowicz, I.A.~Akimov, and M.~Bayer, {\it Access to long-term optical memories using photon echoes retrieved from semiconductor spins}, Nature Photon. \textbf{8}, 851--857 (2014).

\bibitem{Salewski-17} M. Salewski, S.~V. Poltavtsev, I.~A. Yugova, G. Karczewski, M. Wiater, T. Wojtowicz, D.~R. Yakovlev, I.~A. Akimov, T. Meier, and M. Bayer, {\it High-resolution two-dimensional optical spectroscopy of electron spins}, Phys. Rev. X {\bf 7}, 031030 (2017).

\bibitem{Hartmann-69} S. R. Hartmann, {\it Photon, spin, and Raman echoes}, IEEE Journal of Quantum Electronics {\bf 4}, 802 (1968).

\bibitem{Cundiff-91} M. D. Webb, S. T. Cundiff, and D. G. Steel, {\it Stimulated-picosecond-photon-echo studies of localized exciton relaxation and dephasing in GaAs/Al$_x$Ga$_{1-x}$As multiple quantum wells}, Phys. Rev. B {\bf 43}, 12658 (1991).

\bibitem{Poltavtsev-17} S.V. Poltavtsev, M. Reichelt, I. A. Akimov, G. Karczewski, M. Wiater, T. Wojtowicz, D. R. Yakovlev, T. Meier, and M. Bayer, {\it Damping of Rabi oscillations in intensity-dependent photon echoes from exciton complexes in a CdTe/(Cd,Mg)Te single quantum well}, Phys. Rev. B {\bf 96}, 075306 (2017).

\bibitem{Zhukov-07} E. A. Zhukov, D. R. Yakovlev, M. Bayer, M. M. Glazov, E. L. Ivchenko, G. Karczewski, T. Wojtowicz, and J. Kossut, {\it Spin coherence of a two-dimensional electron gas induced by resonant excitation of trions and excitons in CdTe/(Cd,Mg)Te quantum wells}, Phys. Rev. B {\bf 76}, 205310 (2007).

\bibitem{FFT-obzor} S. V. Poltavtsev, I. A. Yugova, I. A. Akimov, D. R. Yakovlev, and M. Bayer, {\it Photon echo from localized excitons in semiconductor nanostructures}, Physics of the Solid State {\bf 60}, 1635 (2018).

\bibitem{Shah-book} J. Shah, {\it Ultrafast Spectroscopy of Semiconductors and Semiconductor Nanostructures}, Vol. 115 of Springer Series in Solid-State Sciences (Springer Science \& Business Media, New York, 1999).

\bibitem{Saeed-18} F. Saeed, M. Kuhnert, I. A. Akimov, V. L. Korenev, G. Karczewski, M. Wiater, T. Wojtowicz, A. Ali, A. S. Bhatti, D. R. Yakovlev, and M. Bayer, {\it Single-beam optical measurement of spin dynamics in CdTe/(Cd,Mg)Te quantum wells}, Phys. Rev. B {\bf 98}, 075308 (2018).

%\bibitem{supplement} See Supplemental Material for the spectral dependence of $g$ factors evaluated from KR, PE and resonant spin amplification methods, and the microscopic theory.

\bibitem{Kavokin-08} K. V. Kavokin, {\it Spin relaxation of localized electrons in $n$-type semiconductors}, Semicond. Sci. Technol. {\bf 23}, 114009 (2008).

\bibitem{Merkulov-02} I. A. Merkulov, Al. L. Efros, and M. Rosen,  {\it Electron spin relaxation by nuclei in semiconductor quantum dots}, Phys. Rev. B {\bf 65}, 205309 (2002).

\bibitem{glazov:hopping} M. M. Glazov, \textit{Spin noise of localized electrons: Interplay of hopping and hyperfine interaction}, Phys. Rev. B {\bf 91}, 195301 (2015).

\bibitem{Kikkawa-98} J.~M. Kikkawa and D.~D. Awschalom, {\it Resonant Spin Amplification in $n$-type GaAs}, Phys. Rev. Lett. {\bf 80}, 4313 (1998).

\bibitem{Sirenko-97} A. A. Sirenko, T. Ruf, and M. Cardona, D. R. Yakovlev, W. Ossau, A. Waag, and G. Landwehr, {\it Electron and hole $g$ factors measured by spin-flip Raman scattering in CdTe/Cd$_{1-x}$Mg$_x$Te single quantum wells}, Phys. Rev. B {\bf 56}, 2114 (1997).

\bibitem{ivchenko05a} E. L. Ivchenko, \emph{Optical Spectroscopy of Semiconductor Nanostructures} (Alpha Science, Harrow UK, 2005).
\bibitem{ES:book} B. I. Shklovskii and A. L. Efros, \emph{Electronic Properties of Doped Semiconductors} (Springer-Verlag, Berlin Heidelberg, 1984).

\bibitem{raikh} R. C. Roundy and M. E. Raikh, \textit{Spin relaxation of a diffusively moving carrier in a random hyperfine field}, Phys. Rev. B {\bf 90}, 201203 (2014).

\bibitem{shumilin} A. V. Shumilin and V. V. Kabanov, \textit{Kinetic equations for hopping transport and spin relaxation in a random magnetic field}, Phys. Rev. B {\bf 92}, 014206 (2015).

\bibitem{rotation1} B. I. Shklovskii, \textit{Dyakonov-Perel spin relaxation near the metal-insulator transition and in hopping transport}, Phys. Rev. B {\bf 73}, 193201 (2006).

\bibitem{rotation2} I.S. Lyubinskiy, A.P. Dmitriev, and V.Yu. Kachorovskii, \textit{Spin dynamics in the regime of hopping conductivity,} JETP Letters {\bf 85}, 55 (2007).




\end{thebibliography}
\end{document}